\documentclass[pre,twocolumn,showpacs,preprintnumbers,superscriptaddress,floatfix]{revtex4}
\usepackage{amsmath,amssymb}
\usepackage{graphicx,epsfig}
\usepackage{dcolumn}
\usepackage{bm}
\begin{document}


\title{Dynamic instability transitions
in 1D driven diffusive flow with nonlocal hopping}

\author{Meesoon Ha}
\affiliation{Department of Physics, Chonbuk National University,
Jeonju 561-756, Korea}
\affiliation{School of Physics, Korea Institute for Advanced Study,
Seoul 130-722, Korea}

\author{Hyunggyu Park}
\affiliation{School of Physics, Korea Institute for Advanced Study,
Seoul 130-722, Korea}

\author{Marcel den Nijs}
\affiliation{Department of Physics, University of Washington,
Seattle, Washington 98195, U.S.A.}

\date{\today}

\begin{abstract}
One-dimensional directed driven stochastic flow with competing
nonlocal and local hopping events has an instability threshold
from a populated phase into an empty-road (ER) phase. We implement
this in the context of the asymmetric exclusion process. The
nonlocal skids promote strong clustering in the stationary
populated phase. Such clusters drive the dynamic phase transition
and determine its scaling properties. We numerically establish
that the instability transition into the ER phase is second-order
in the regime where the entry point reservoir controls the current
and first order in the regime where the bulk is in control. The
first order transition originates from a turn-about of the cluster
drift velocity. At the critical line, the current remains
analytic, the road density vanishes linearly, and fluctuations
scale as uncorrelated noise. A self-consistent cluster dynamics
analysis explains why these scaling properties remain that simple.

\end{abstract}


\maketitle

\narrowtext

\section{Introduction}
\label{intro}

Driven stochastic diffusive flow of particles in narrow channels
is widely used as a prototype process to study dynamic scaling.
The central interests in such processes reside with the structure
of non-equilibrium stationary (NES) states, the dynamical pathways
to those NES states, and the scaling properties of dynamic phase
transitions. Driven stochastic processes often undergo dynamic
phase transitions inside their NES states as a function of control
parameters. This is true even in one dimension where equilibrium
phase transitions are forbidden. One-dimensional (1D) processes
serve thus as ideal platforms to unearth novel general principles
affecting the dynamic scaling properties and the structure of NES
state distributions. The maximum entropy principle familiar from
thermal equilibrium, leading to the Gibbs distribution, does not
apply to driven systems. We need to formulate new principles, if
possible, to predict the structures of NES
states~\cite{SZia-DL-v17,Schuetz-DL-v19}. Such principles can
emerge from the study of a wide array of specific processes, in
particular using numerical simulations and/or exact solutions.

In this paper, we address how clustering induced by nonlocal
stick-slip events affects 1D driven stochastic diffusive flow. Our
model is a generalization of the conventional asymmetric
exclusion process (ASEP)~\cite{MacDonald,Derrida}. We allow
nonlocal forward hopping events across empty stretches of road,
which introduces an instability towards an `empty-road' (ER) phase.

Clustering and queuing has been studied recently in different
generalizations of ASEP. In the conventional ASEP, particles hop
stochastically in a preferred direction along a loop, without being able
to pass each other and with the occupation of each site limited to one.
That ASEP stationary state is surprisingly trivial, completely
disordered. However, it is
very sensitive to defects and boundary conditions.
The sensitivity to reservoirs in open boundary type set-ups was
discovered first. In that set-up the stationary state is known
exactly by the so-called matrix product ansatz~\cite{Derrida, Schuetz-DL-v19}.
The two reservoirs at the edges of the channel compete with the bulk
for control over the bulk density $\rho$ and the average flow rate.
This gives rise to phase transitions between bulk and
reservoir controlled phases. The latter can be viewed as
elementary forms of queuing (traffic jams). The density profiles
near reservoirs have exponential or power-law tails. The scaling
properties remain rather simple and are qualitatively understood from
the fact that the ASEP bulk stationary state is fully uncorrelated,
while fluctuations spread as $l\sim t^{1/z}$ with $z=3/2$, and travel
with group velocity $v_{g}=1-2\rho $ (see e.g., Ref.~\cite{PGM}).

Introducing bulk impurities to ASEP, e.g., a single slow-bond
in the bulk, represents the next level of complexity in queuing.
Such a single static defect creates more intricate types of
jamming than edges, because, unlike reservoirs,
information and correlations can travel across the obstacle.
Recently we demonstrated that the scaling properties of the
slow-bond queuing transition are non-trivial. We established the
existence of an intermediate phase with a pure power-law shaped
queue, and showed that the transition into the macroscopic jammed
phase does not take place until a finite strength of the
obstruction~\cite{SB-exp,SB-theory}.

A next step in the queuing and clustering saga is to allow such
impurities to be mobile and participate in the dynamics.
Macroscopic clustering represents jamming. For example, in the two species
process by Arndt {\it et al.}~\cite{AHR} particles of opposite charge hop
in opposite directions and the passing of species
creates jamming. The stationary states of such processes show
strong clustering. The possibility of phase separation into
macroscopic clustered states is discussed at length in the literature
both numerically~\cite{AHR-wrong} and analytically~\cite{Mukamel}
as well as the link to the zero-range process (ZRP)~\cite{ZRP}.

In this paper, we study queuing caused by a different mechanism.
In our generalization of ASEP,
clustering is induced by nonlocal hopping events. A particle can
jump with probability $1-p$ to the nearest-neighbor site (if
empty) or with probability $p$ all the way forward to the site
immediately behind the particle in front of it.
We use open boundary conditions with reservoirs at both edges.
Starting-off as uniform at $p=0$, the stationary state becomes
increasingly lumpy with well-defined clusters as $p$ increases.
Then, at a critical value $p_c$, this clustered liquid becomes
unstable and the road empties out. The location and the nature of
this phase transition are our main concerns. The transition turns
out to be discontinuous or critical. Clustering and coarsening
prove to be essential in explaining both.

Clustering is quite common in driven stochastic flow. It
appeared for example in the two species type process by Arndt {\it
et al}.~\cite{AHR}. Those clusters were successfully described by
the ZRP~\cite{ZRP}, which have zero drift. In our process, the
nonlocality induces clustering with nonzero drift. More
importantly, that aspect induces first- and second-order phase
transitions, and the scaling properties of those are governed by
the clustering.

This process and its phase transitions also relate to various
viscous and/or dissipative stick-slip type phenomena, where the
particle number is the only locally conserved variable. Potential
applications include traffic and/or granular flow~\cite{traffic},
the motion of stick and flow in sandpiles~\cite{sandpile}, phase
separation in steady sedimentation of colloidal
crystals~\cite{colloid-theory,colloid-exp}, electronic and/or
molecular transport in nanoscale systems~\cite{transport}, phase
ordering in rough films~\cite{DKardar}, the motion of molecular
motors driven by ATP~\cite{motors}, the motion of a depinned flux
lattice in a current-carrying superconductor~\cite{superconductor}
and so on. Nonlocal hopping can be used to mimic the dynamic features
associated with, e.g., the competition between maximum and minimum
speed/drift in traffic/granular flow, inertia of falling grains
inside avalanche processes in sandpiles, and the role of gravity
in dynamic and static sedimentation. 

This paper is organized as follows. In section~\ref{model}, we
present numerical results for the phase diagram, clustering,
and the nature of the phase transitions. In section~\ref{free},
we setup a self-consistent free cluster analysis, in which the
stationary state is described in terms of a collection of freely
drifting clusters, i.e., self-organized mesoscopic collective
objects that absorb and emit individual particles constantly. The
first-order segment of the phase transition line is explained as the
result of a turn-about in the drift velocity of those free clusters.
The critical segment of the transition line is caused by starvation and
the fluctuations in the density near the entry reservoir become crucial.
To describe this, we extend in section~\ref{mother}
the self-consistent cluster analysis to the mother cluster, which
is the cluster attached to the  entry point reservoir.
This analysis explains the numerical details of the
critical transition presented in section~\ref{numerical}, and discussed in
section~\ref{separate}. The paper concludes, in section~\ref{ending},
with a brief summary of the results and some open questions.

Our self-consistent cluster approximations resemble recent
``zero-range process" (ZRP) treatments of clustering phenomena, e.g.,
in the ASEP with two or more species of particles (driven in opposite
directions)~\cite{Mukamel}. Our approach is both more basic and
more general. In those processes the clusters are stationary on
average, while ours have a non-zero drift velocity.
In the ZRP approach the internal structure of the clusters
is ignored, while in our approach the internal density profile is
taken into account.
Both aspects are essential for the transition into the ER phase.
ZRP can be solved exactly,
while our discussion remains more approximative.

\section{Phase diagram}
\label{model}

\begin{figure}[t]
\includegraphics*[width=0.7\columnwidth, angle=270]{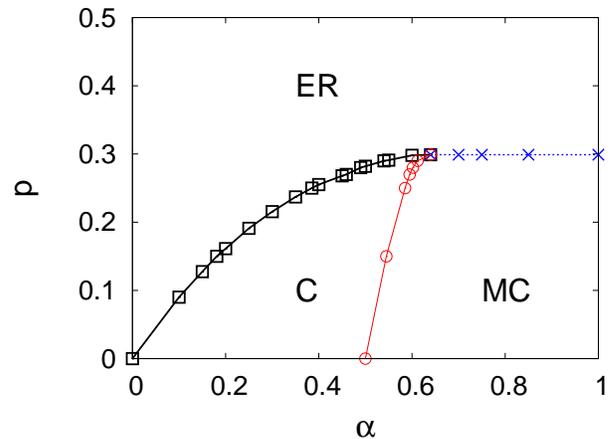}
\caption{(Color online)~$\alpha-p$ phase diagram from our numerics.}
\label{fig.palpha}
\end{figure}

Consider a chain of length $1\leq x\leq L$ with sites that can
only be empty or occupied, $n_x=0,1$. 
The chain is open, in contact with the entry side reservoir
at $x=0$ and the exit side reservoir at $x=L+1$. The updated rule
for our process is as follows:

(1)Select a site, $0\le x\le L$, at random.

(2)If $x\ne 0$
and it is occupied, the particle slides to its nearest neighbor
site $x+1$ (with probability $1-p$) or detaches and jumps all the
way to the empty site directly behind the nearest particle in
front of it (with probability $p$). 

(3)In case the chain is
completely empty in front of it, the particle jumps all the way
forward into the exit side reservoir.

(4)If the entry side
reservoir, $x=0$, is selected and site $x=1$ is empty, a particle
jumps onto site $x=1$ with probability $\alpha$.
Nonlocal hopping events from the reservoir are not allowed in our
set-up.

\begin{figure*}[ht]
{\bf (a)
$p=0 \hspace*{1.5cm}
0.1\hspace*{1.75cm}0.15\hspace*{2cm}0.2\hspace*{1.75cm}0.25\hspace*{1.5cm}0.3$}

\vspace{-0.25cm}
\begin{center}
\begin{tabular}{cccccc}
\includegraphics*[width=0.25\columnwidth, angle=0]{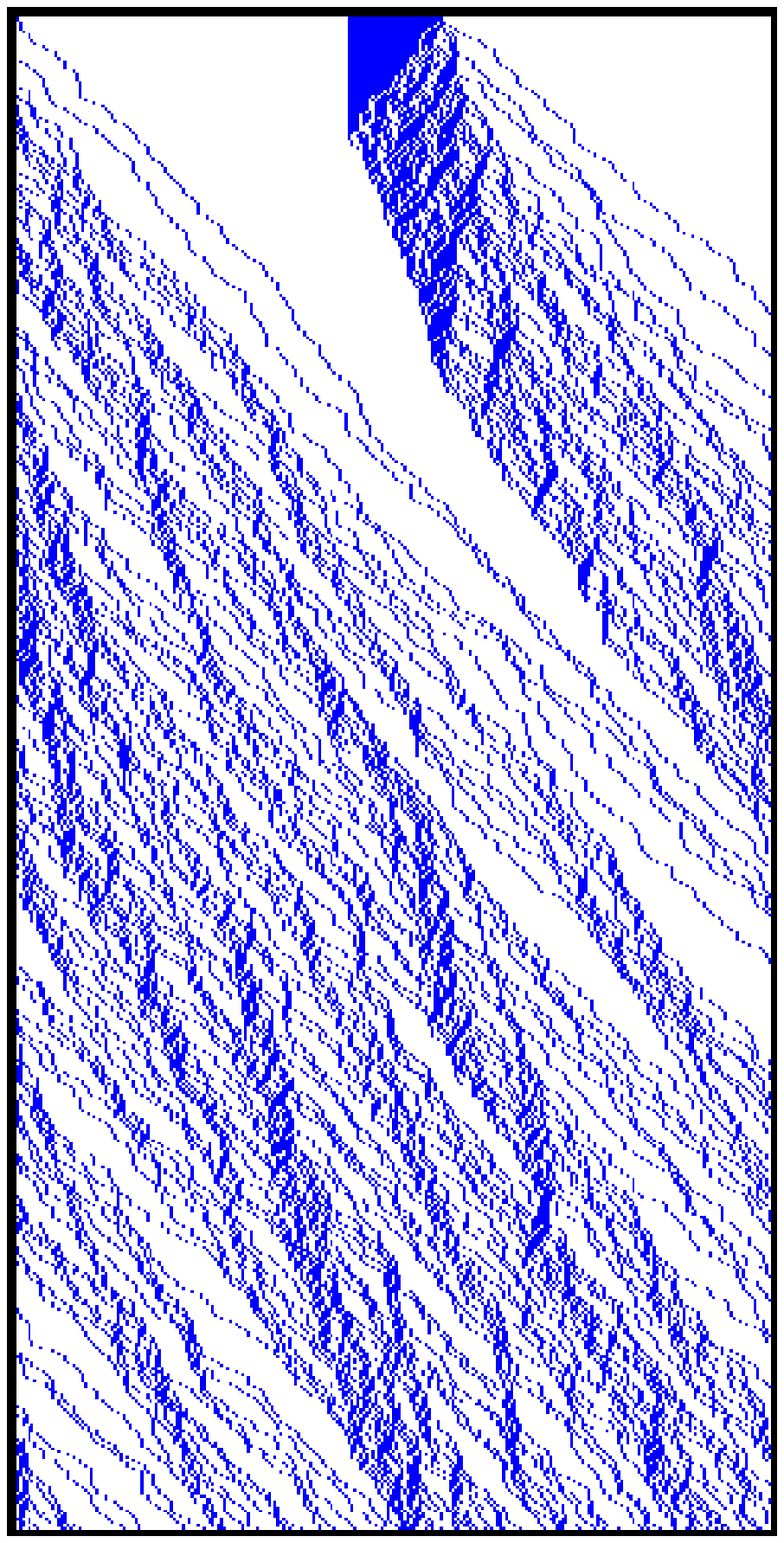}&
\includegraphics*[width=0.25\columnwidth, angle=0]{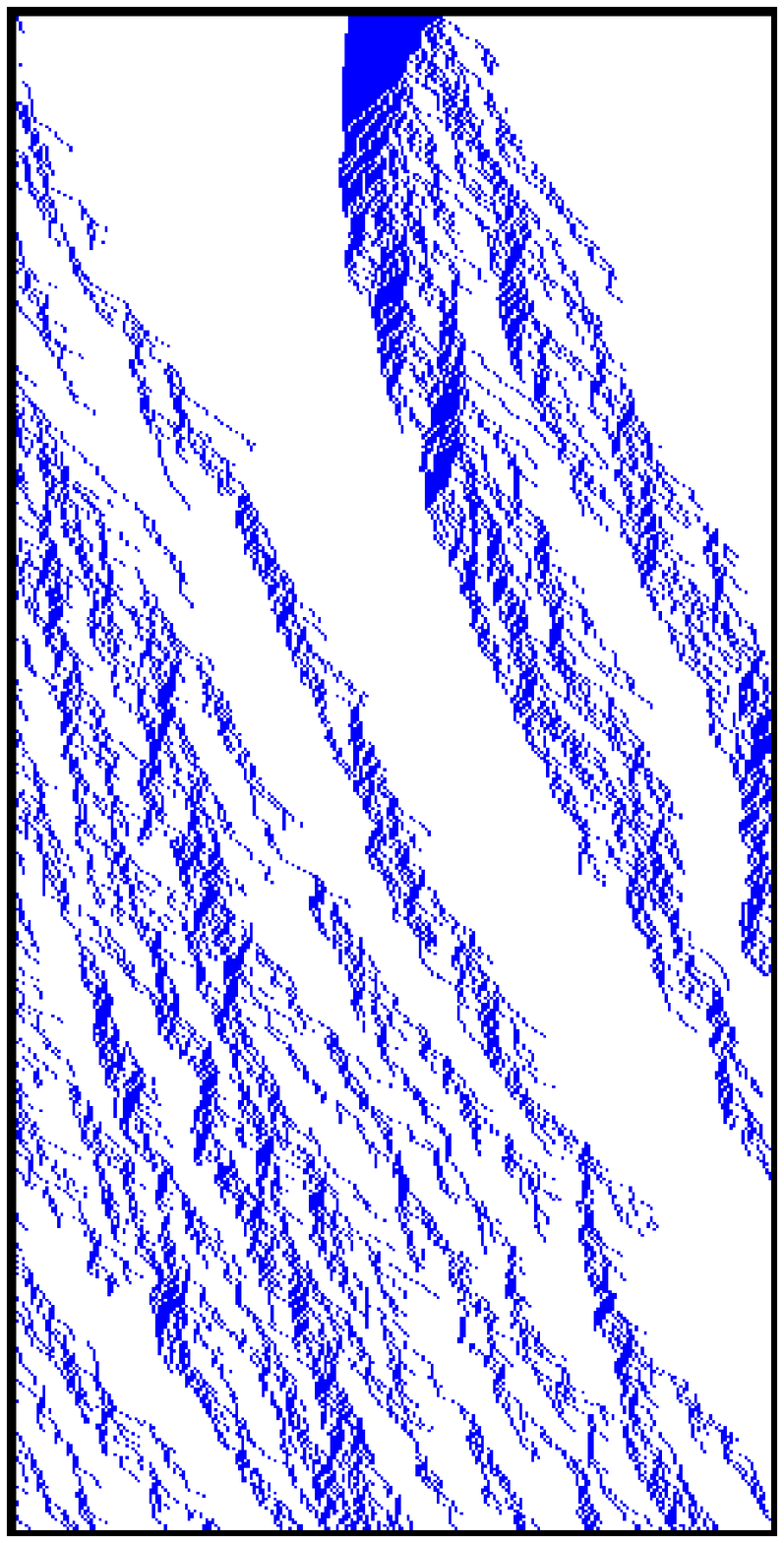}&
\includegraphics*[width=0.25\columnwidth, angle=0]{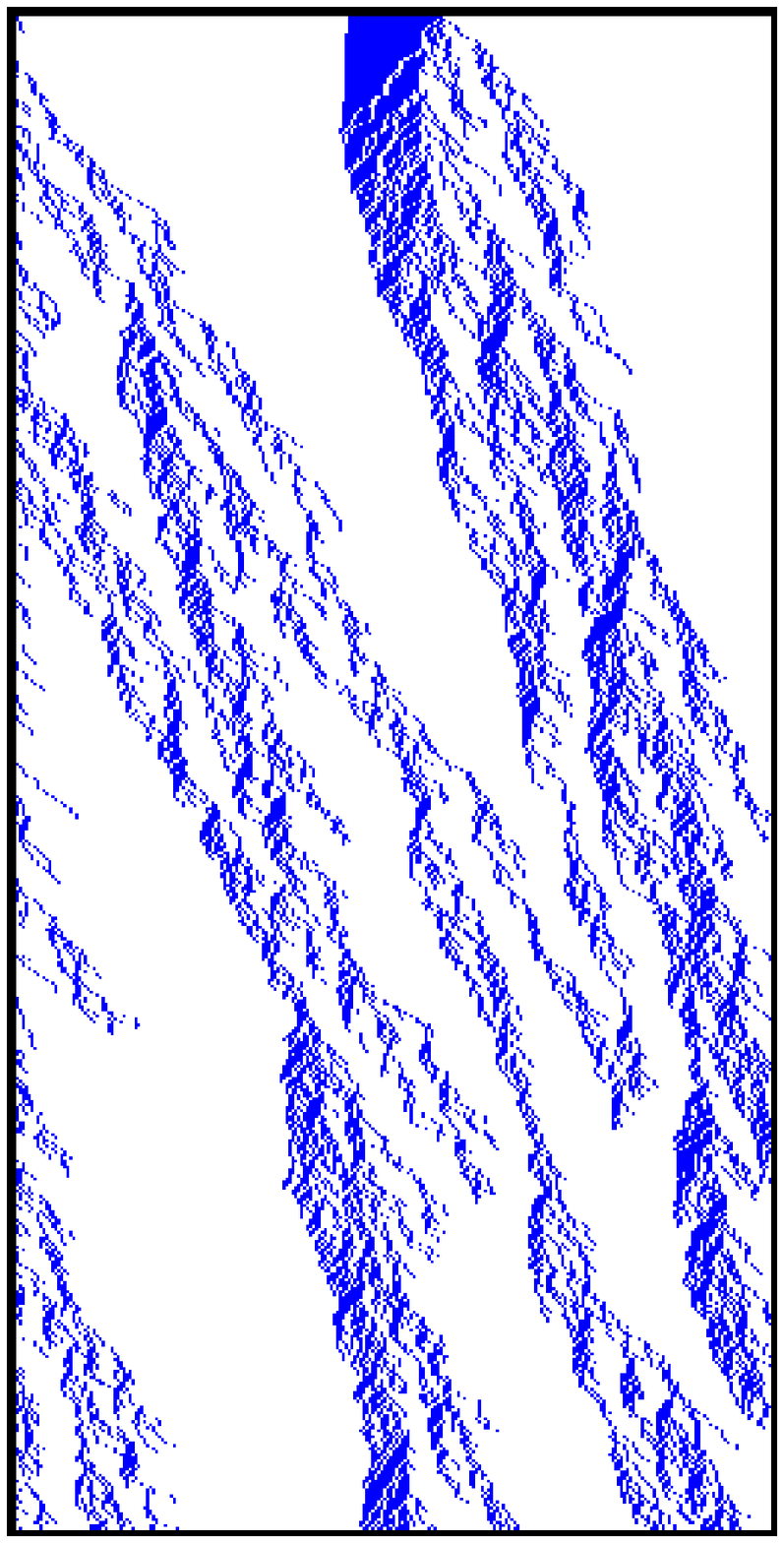}&
\includegraphics*[width=0.25\columnwidth, angle=0]{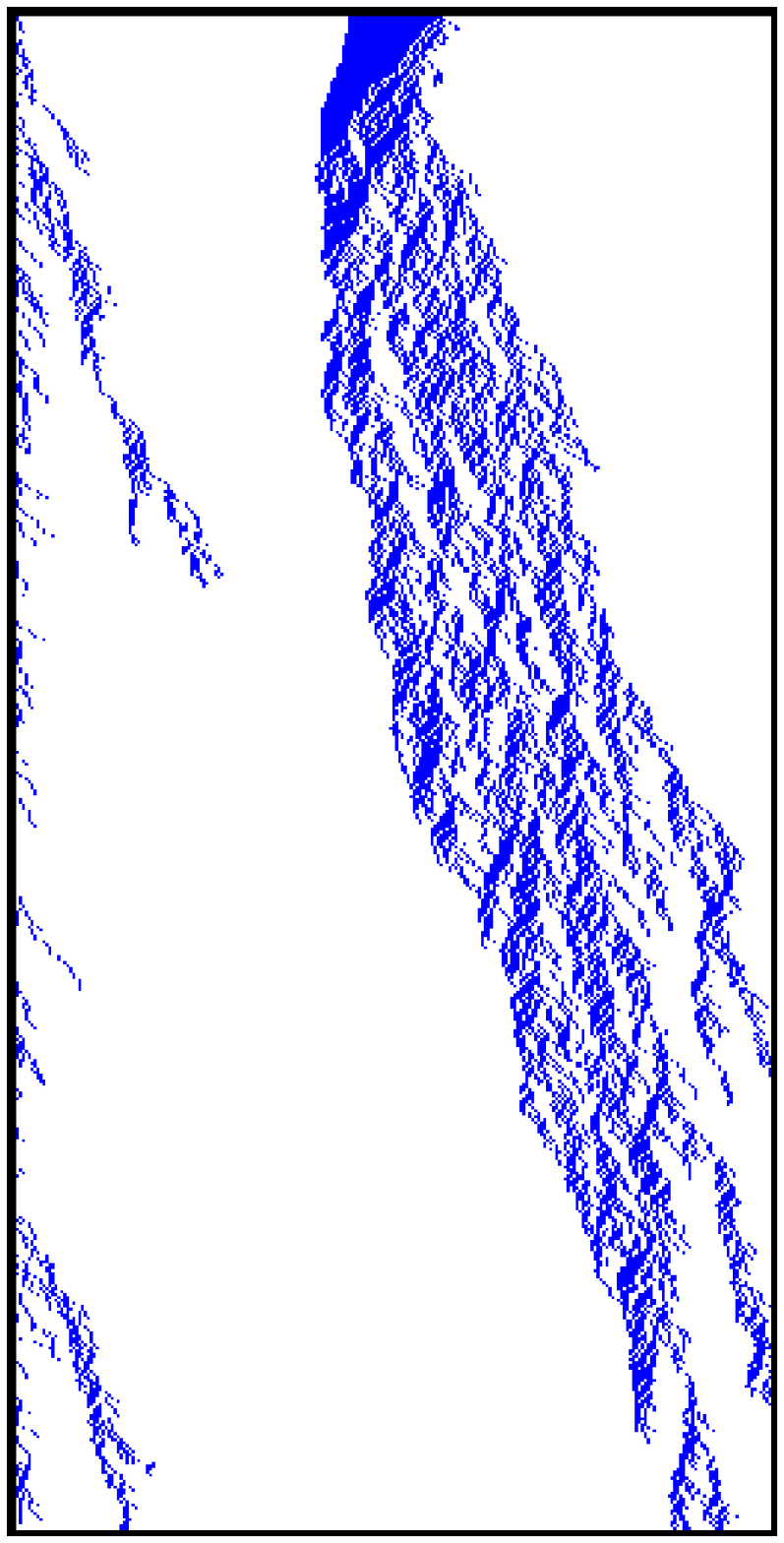}&
\includegraphics*[width=0.25\columnwidth, angle=0]{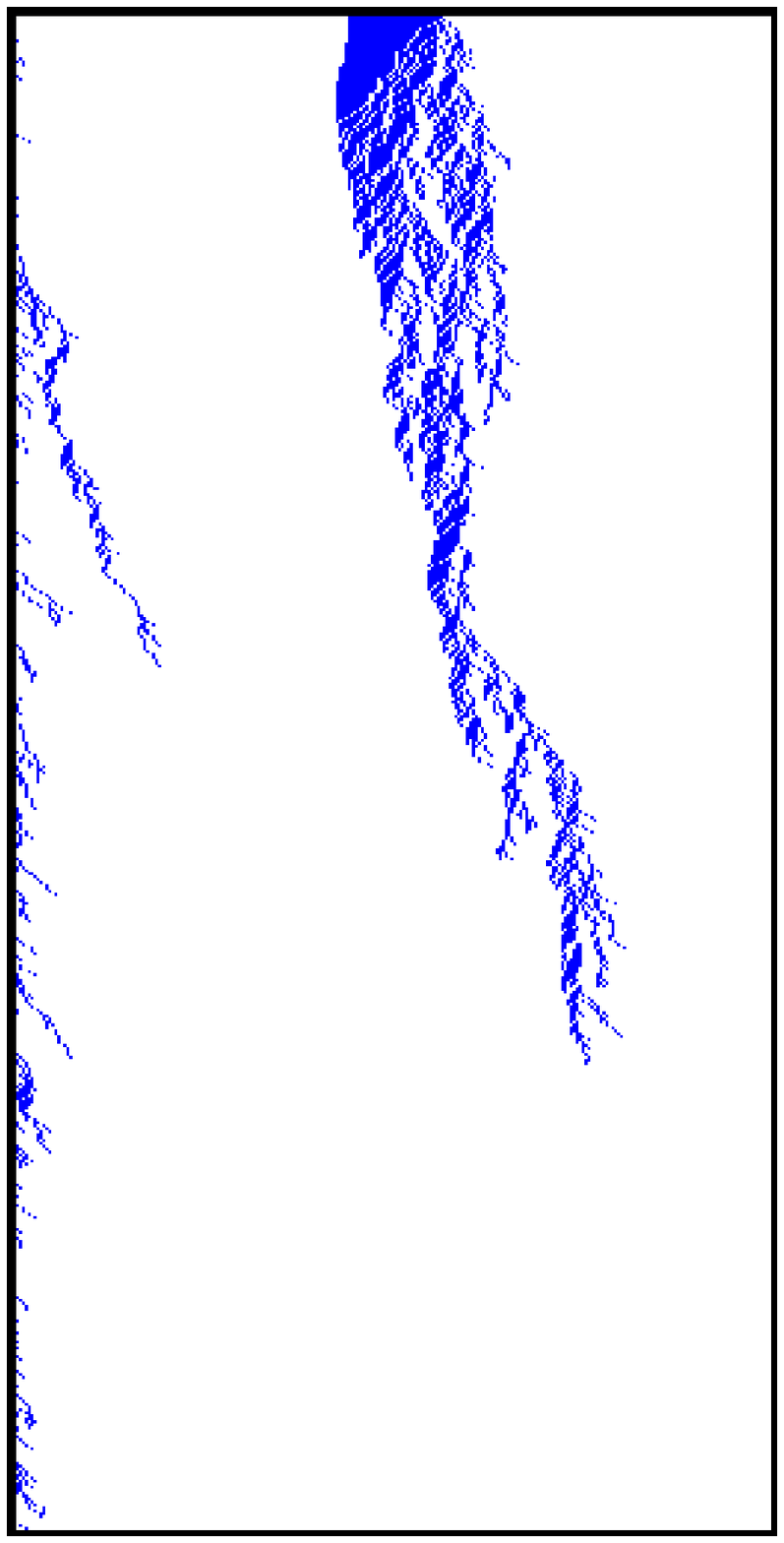}&
\includegraphics*[width=0.25\columnwidth, angle=0]{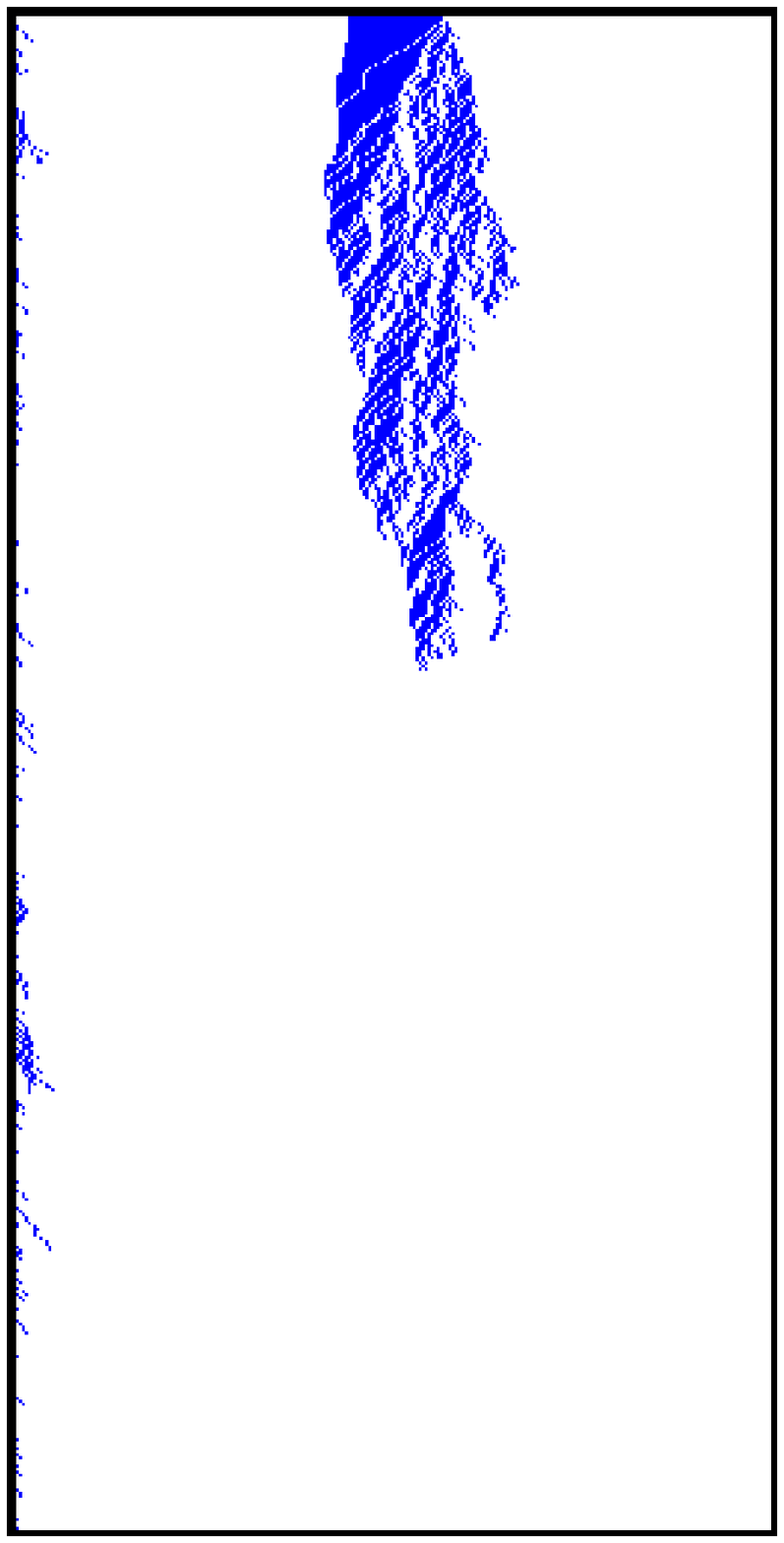}
\end{tabular}
\end{center}

{\bf (b)
$p=0 \hspace*{1.75cm}
0.1\hspace*{2cm}0.2\hspace*{2cm}0.3\hspace*{1.75cm}0.4\hspace*{1.75cm}0.5$}

\vspace{-0.25cm}
\begin{center}
\begin{tabular}{cccccc}
\includegraphics*[width=0.25\columnwidth, angle=0]{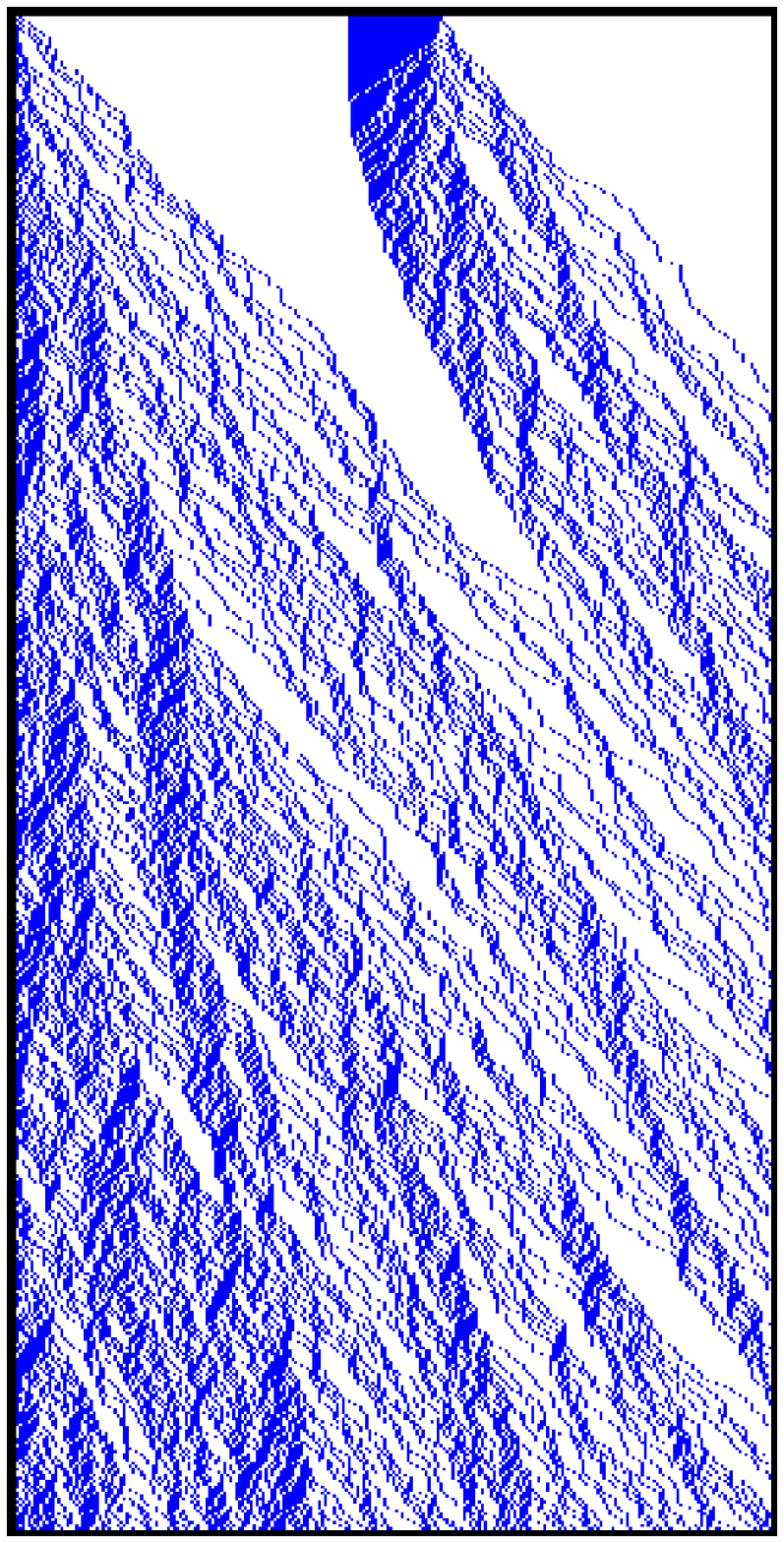}&
\includegraphics*[width=0.25\columnwidth, angle=0]{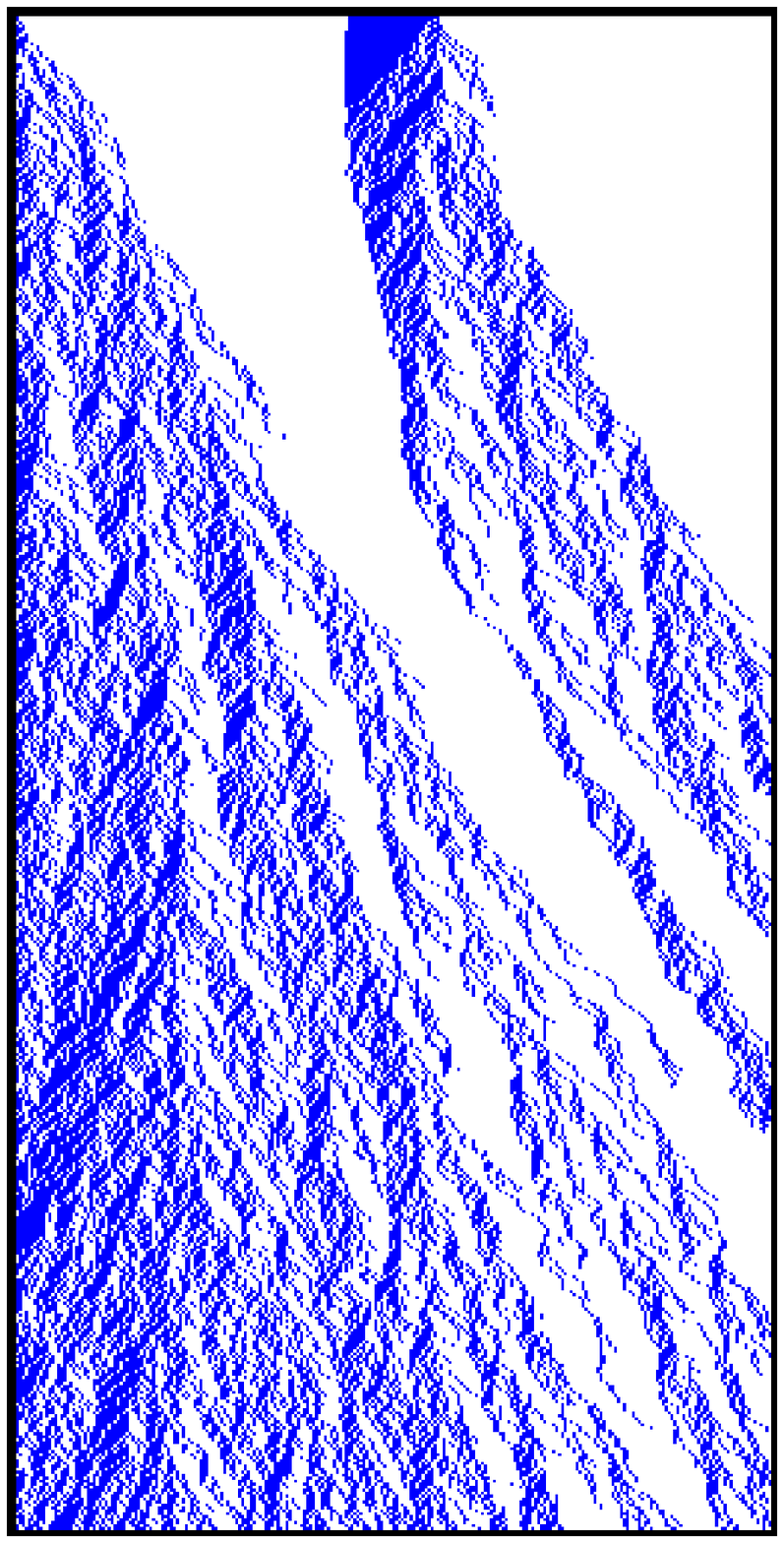}&
\includegraphics*[width=0.25\columnwidth, angle=0]{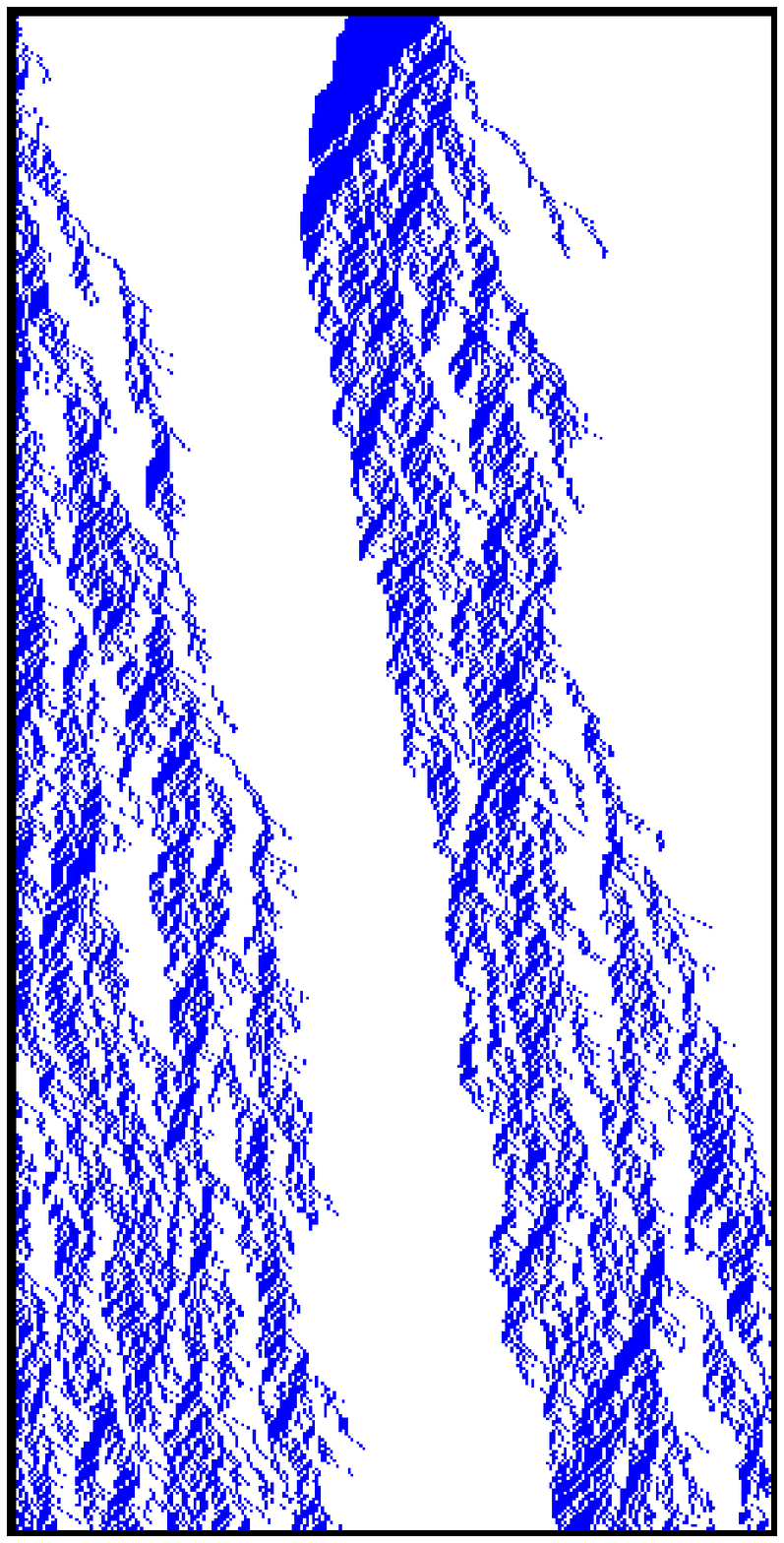}&
\includegraphics*[width=0.25\columnwidth, angle=0]{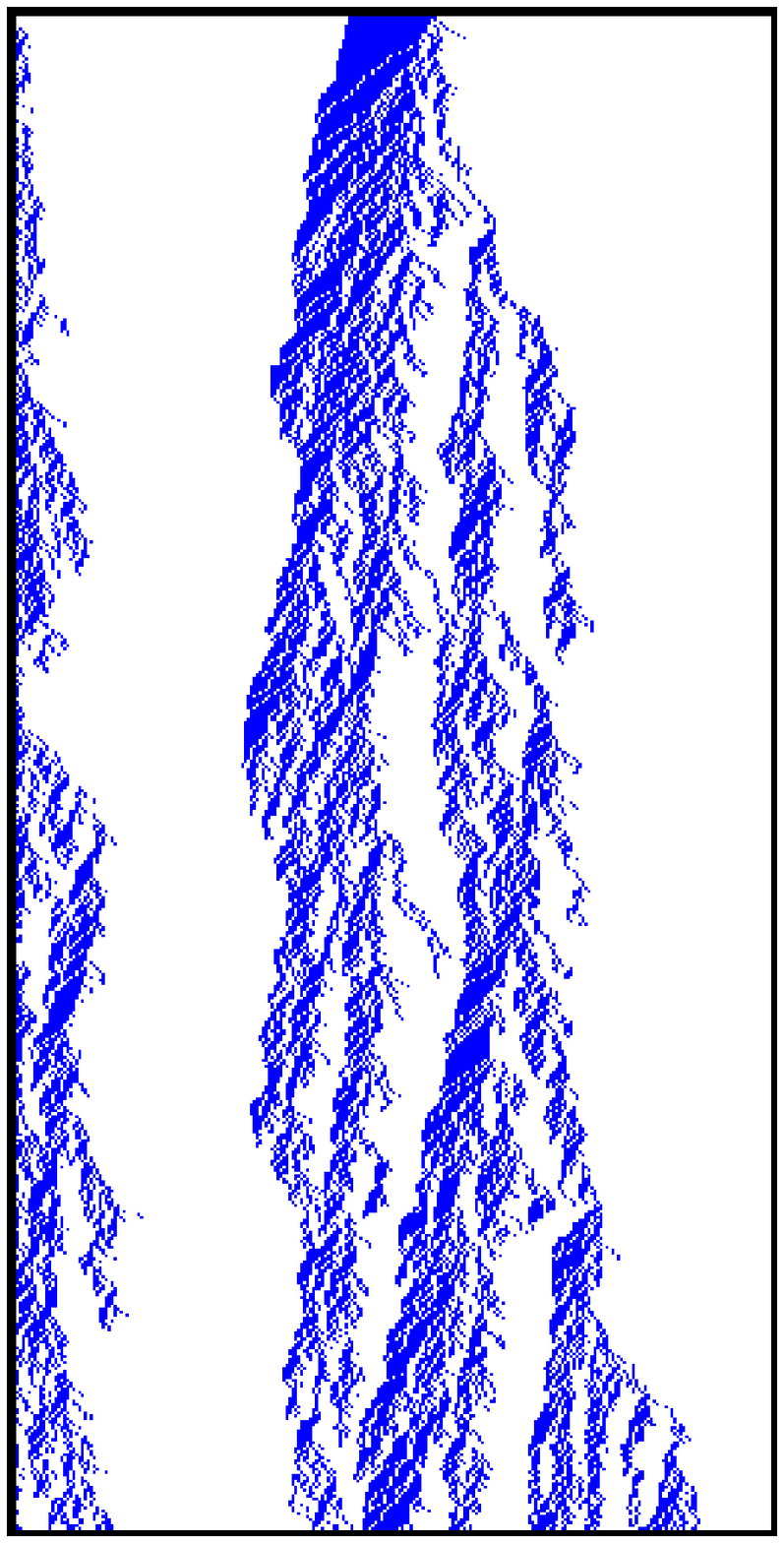}&
\includegraphics*[width=0.25\columnwidth, angle=0]{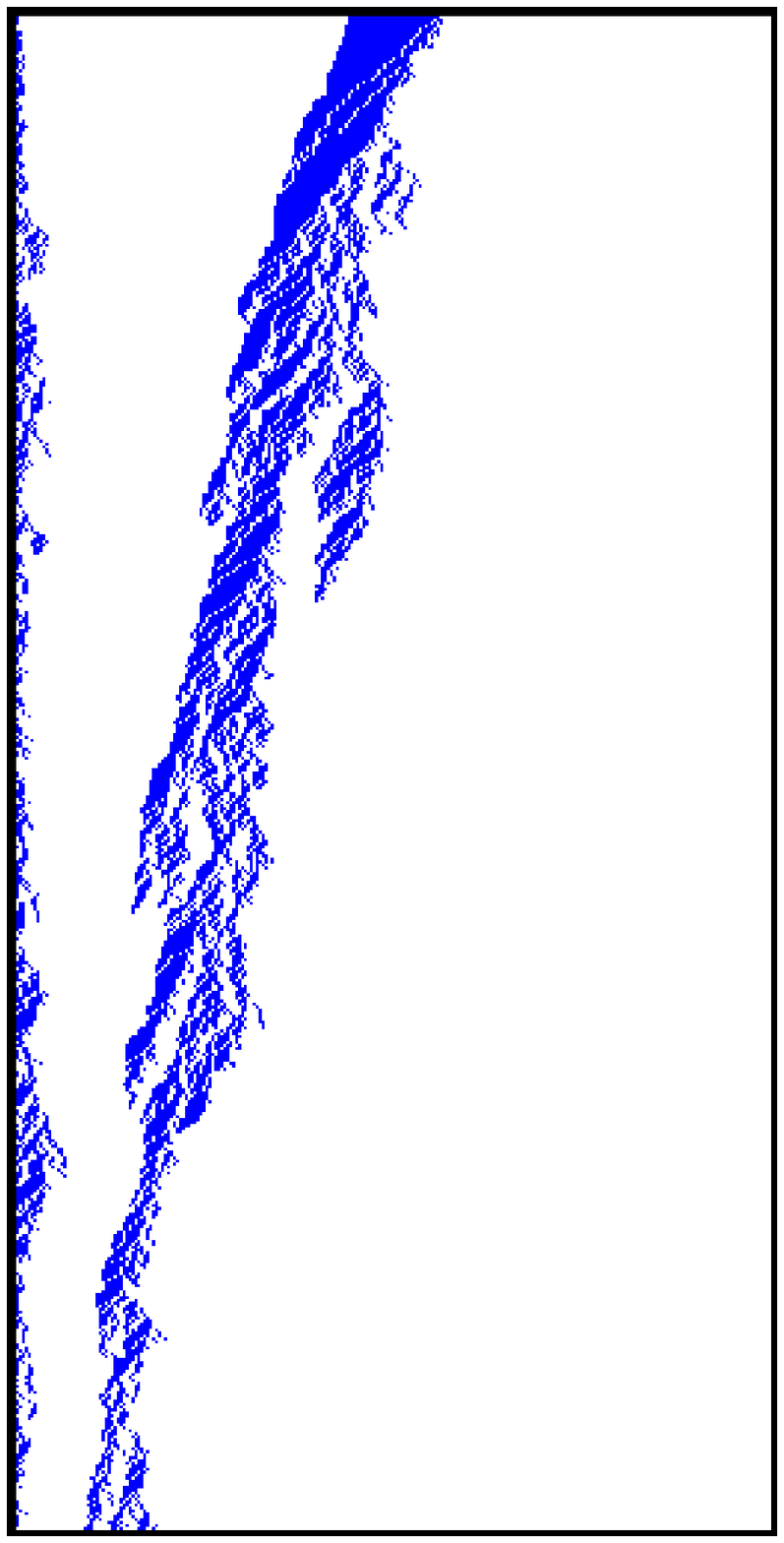}&
\includegraphics*[width=0.25\columnwidth, angle=0]{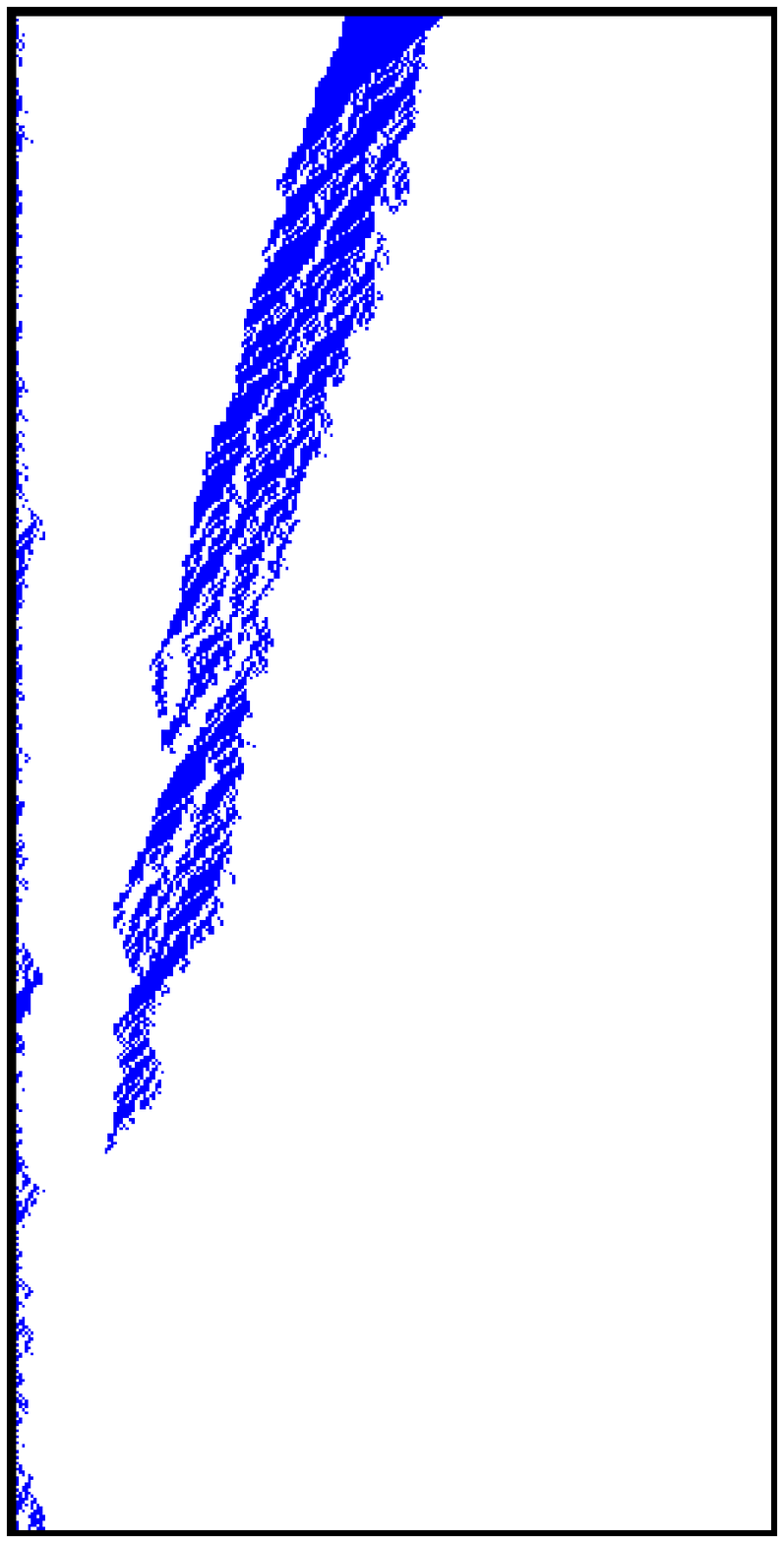}
\end{tabular}
\end{center}
\caption{(Color online) Spatiotemporal density profiles for $L=256$ 
($1\le x\le$ running horizontally from left to right)
up to $t=512$ ($0\le t\le$ running vertically from top to bottom)
illustrate clustering in both (a) ($\alpha=0.25:~p_c\simeq 0.19$) the C and
(b) ($alpha=1.0:~ p_c\simeq 0.30$) the MC phase. 
Initial conditions include one compact
free cluster in the middle of the system to show clearly
that the drift velocity of
the free clusters turns around at $p\simeq 0.3$.}
\label{fig.clustering}
\end{figure*}

Figure~\ref{fig.palpha} shows the phase diagram obtained from our
Monte Carlo simulations with road lengths up to $L=16~000$. At
$p=0$ the process reduces to the conventional ASEP. There, the
stationary state is well-known to be in either: the input
reservoir-controlled phase (C) for $\alpha<\frac{1}{2}$, where the
bulk density is set and limited  by the influx fugacity $\alpha$;
or the bulk-controlled phase (MC) for $\alpha>\frac{1}{2}$, where
the influx rate is not the limiting factor anymore and the bulk
density is set by maximizing the bulk current.

The C and MC phases persist into $p>0$, but become increasingly
clustered, see Fig.~\ref{fig.clustering}. Their historical names
become quite unfortunate. `MC phase' stands in the
conventional ASEP context, for `maximal current phase'. Control
over the current, by either the bulk or the reservoir, is closely
linked at $p=0$ with the so-called `maximal current principle'.
The latter has been proposed as a general principle to predict
dynamic stationary states. Its success in the conventional ASEP is
closely tied to the stationary state being uniform and
clusterless, and well-approximated by mean-field
(MF) theory. The maximal current principle arose naturally as a
reformulation of the MF iteration process. Its generalization to
strongly clustered states remains unclear, and this issue plays an
important role in the discussion below.

\begin{figure}[t]
\includegraphics*[width=0.7\columnwidth, angle=0]{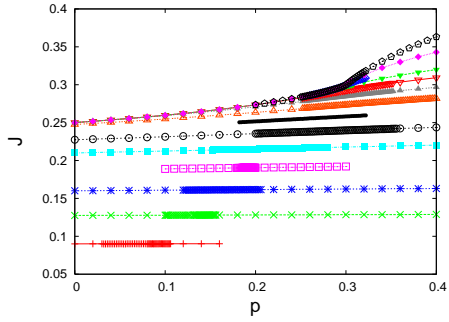}
\caption{(Color online) The steady-state current $J$ versus $p$
for $L=8000$ at fixed $\alpha=0.1, 0.15,\cdots,1$ from bottom to
top (with the same $\alpha$-labelling as in
Fig.~\ref{fig.density-p}).} \label{fig.current-p}
\end{figure}

`Bulk-in-Control' and
`Reservoir-Controlled' would be better names for the two
phases in Fig.~\ref{fig.clustering}, but we avoid confusion by
preserving their historical abbreviations `MC'
(`the-Middle-in-Control' phase) and `C'
(`being-Controlled'-phase). In the C phase, the stationary-state
current varies both with $p$ and $\alpha$, while in the MC phase
it only varies with $p$ regardless of boundary conditions,
i.e., $\alpha$-independent; see Fig.~\ref{fig.current-p}.
This feature distinguishes the two phases in our numerical analysis.

\begin{figure}[b]
\includegraphics*[width=0.7\columnwidth, angle=0]{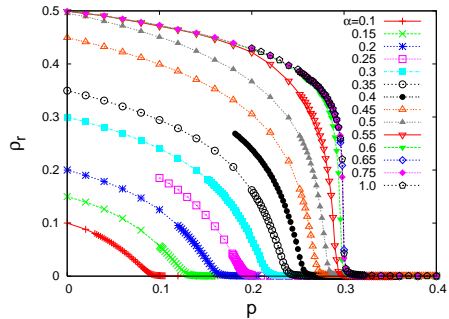}
\caption{(Color online) The steady-state bulk road particle density
$\rho_r$
versus $p$ at fixed $\alpha=0.1, 0.15,\cdots,1.0$ for $L=8000$.}
\label{fig.density-p}
\end{figure}

\begin{figure}[t]
\includegraphics*[width=0.7\columnwidth, angle=0]{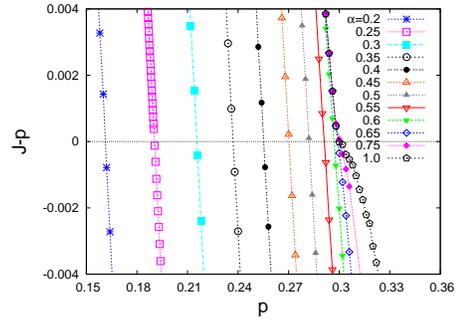}
\caption{(Color online) $J-p$ versus $p$ at fixed $\alpha=0.2,
0.25,\cdots,1.0$
for $L=8000$. The intersection points with the $J-p=0$ line
represent the C-ER and the MC-ER phase transition points.}
\label{fig.current-p-p}
\end{figure}

For strong nonlocal hopping (large $p$), the stationary state
transforms into an `empty road' (ER) phase. The transition line
starts at small $\alpha$ as a critical line, located just below
$p=\alpha$. It levels-off for increasing $\alpha$, and changes at
the critical endpoint, at $\alpha=0.64(1)$, into a first-order
transition line. The latter is a strictly horizontal line, with $p_c= 0.300(2)$,
see Fig.~\ref{fig.palpha}. The density of particles
$\rho_r$ jumps to zero at the first-order line, see
Fig.~\ref{fig.density-p}, but the average current remains
continuous, see Figs.~\ref{fig.current-p} and~\ref{fig.current-p-p}.

\begin{figure}[b]
\includegraphics*[width=0.7\columnwidth, angle=0] {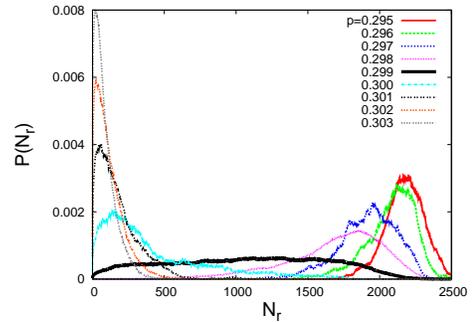}
\caption{(Color online) The probability distribution $P(N_r)$ for
finding $N_r$ particles on the road, at $\alpha=0.75$ for various
$p=0.295,~0.296,...,~0.303$ (from right to left) at $L=8000$. The
MC-ER transition occurs at $p_c \approx 0.299$. }
\label{fig.histoMC}
\end{figure}

\begin{figure}[t]
\includegraphics*[width=0.7\columnwidth, angle=0]{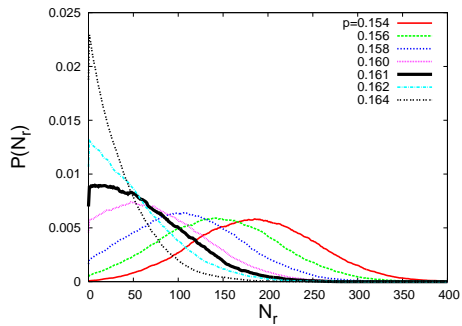}
\caption{(Color online) The probability distribution $P(N_r)$ for
finding $N_r$ particles on the road, at $\alpha=0.2$ for various
$p=0.154,~0.156,...,~0.164$ (from right to left) at $L=8000$. The
C-ER transition occurs at $p_c \approx 0.161$.} \label{fig.histoC}
\end{figure}

The road is not truly empty in the ER phase. The average bulk density vanishes,
but typically a finite cluster of particles `hangs' still from the entry side reservoir
near $x=1$,
and some isolated clusters and individual particles are traveling
through the bulk. Figs.~\ref{fig.histoMC} and \ref{fig.histoC} illustrate this.
They show the evolution of the probability distribution for
finding $N_r=\rho_r L$ particles on the road across the
first-order transition, at $\alpha=0.75$, and across the
critical line, at $\alpha=0.2$. Note the gradual evolution of the
distribution across $p_c$ at $\alpha=0.2$ versus the abrupt change
at $\alpha=0.75$.

Consider the average current through a bulk bond between $x$ and $x+1$
\begin{eqnarray}
\langle J_{x+\frac{1}{2}} \rangle = \langle n_x v_{x+1} \rangle +
p\langle v_x v_{x+1}\rangle -pP_0(x+1),
\label{bulkcurrent}
\end{eqnarray}
and through the edge bonds
\begin{eqnarray}
\label{edgecurrent}
\langle J_{\frac{1}{2}} \rangle =
\alpha\langle v_1 \rangle ~, ~ \langle J_{L+\frac{1}{2}} \rangle =
\langle n_{L}\rangle +p \langle v_{L}\rangle -pP_0(L),
\end{eqnarray}
where $n_x$ is the occupation operator at site $x$, and $v_x\equiv
1-n_x$ is the vacancy operator. Equation~(\ref{bulkcurrent}) states
that the current through the bulk bond is zero if site $x+1$ is
occupied; equal to one if site $x+1$ is empty while
$x$ is occupied; and equal to $p$ if both sites are
empty provided at least one particle exists on the road somewhere
to the left of $x+1$. The latter requires the introduction of the
vacancy string operator
\begin{eqnarray}
\label{emptyprob} P_0(x)= \langle \prod_{y=1}^{x} v_y\rangle,
\end{eqnarray}
which counts the probability for the entire road to the left of
$x+1$ being empty. $P_0(L)$ acts as an order parameter, which
vanishes in the C and MC phases, but remains
non-zero across the entire chain in the ER phase.
$P_0(x)$ is always non-zero near the entrance at small $x$.


The qualitative structure of the phase diagram is now easily understood.
In the stationary state, the average current is constant throughout
the system; and Eq.~(\ref{edgecurrent}) implies the identity
\begin{eqnarray}
\label{edgecurrentcondition}
\alpha \langle v_1 \rangle =
1-(1-p)\langle v_L\rangle-pP_0(L)
\end{eqnarray}
This excludes the C and MC phases from extending into the $p>\alpha$
region of the phase diagram. The identity cannot be satisfied for
$p>\alpha$ with $P_0(L)=0$; the right hand side varies between $p$ and 1,
because $0\leq v_{L}\leq 1$,
and that is inconsistent with $0\leq v_1\leq 1$ for $p>\alpha$.

Equation~(\ref{edgecurrent}) also implies that the current along the
entire transition line is equal to $J=p_c(\alpha)$; because at
$p_c$, $\langle v_{L}\rangle$ is already equal to one, while
$P_0(L)$ is still equal to zero. Our numerical simulations confirm
this.

The levelling-off of the transition line into a horizontal
(constant $p$) segment is linked to this as well. As a start,
we consider MF theory. In the MC phase, where the bulk controls
the current, Eq.~(\ref{bulkcurrent}) is then approximated as
$J=v_b-(1-p)v_b^2$ where $v_b$ is the vacancy density deep inside
the bulk. The maximum current representation of MF theory,
${\delta J}/{\delta v_b}=0$, yields, $v_b= \frac{1}{2r}$ and $J=
\frac{1}{4r}$, with $r=1-p$. This together with the condition
$J(p_c)=p_c$ puts the MC-ER transition line at $p=\frac{1}{2}$.

This is only an upper limit to its true location. The clustering
in the MC phase, shown in Fig.~\ref{fig.clustering}, is not
represented in MF theory. It underestimates the number of
nearest neighbor pairs. These pairs shift the transition line
downwards, see Eq.~(\ref{bulkcurrent}). More advanced versions of
MF theory incorporate local correlations but still fail to
account for clustering. Instead, we develop in the following
section a self-consistent cluster approach. It predicts the
correct location of the MC-ER transition line, and more
importantly, it provides insight in the mechanism
that makes the MC-ER transition discontinuous.

We note that a closed chain (with periodic boundary
conditions) does not include the ER phase due to conservation of
the total number of particles and holes in such a set-up. A
similar model was studied in the context of the mass-conserving
coalescence process, the so-called mass-chipping
model~\cite{mass2,mass3} with and without a bias. Implications of
our results in this context will be discussed
elsewhere~\cite{OurStudy}. 

\section{Free cluster analysis}
\label{free}

Adopt the point of view where the C and MC type stationary state
configurations are fully clustered; with one distinct `mother
cluster' attached to the entry reservoir at small $x$ and many
disconnected freely drifting clusters in the bulk. These clusters
are meant to be only mesoscopic in size, because true phase separation in
terms of macroscopic high density (clusters) and low density
sections is not realized in this process, see
Fig.~\ref{fig.clustering}.

Assume the bulk clusters are compact non-fractal mesoscopic
objects with average bulk density $\rho_c$ and local average
densities, $\rho_F$ and $\rho_R$, at the front $x_F$ and rear $x_R$
end, respectively. The front moves forward by creep (local hops,
$x_F\to x_F+1$) and backward by detachment of the front particle
by a nonlocal hop to the next cluster, with velocity
\begin{eqnarray}
\label{clusterfront}
u_F= \frac{\partial x_F }{\partial  t} =(1-p)
-p/\rho_F
\end{eqnarray}
The rear moves by local forward hops (internal nonlocal hops are
suppressed by the high particle density inside the clusters) and
by attachment of a new particle due to a nonlocal hop from the
free cluster behind it.
\begin{eqnarray}
\label{freeclusterrear}
u_R= \frac{\partial x_R }{\partial  t} =
(1-\rho_R) -p
\end{eqnarray}
The drift velocity $u_d$ of the cluster is associated with its center of mass
\begin{eqnarray}
\label{freeclusterdrift} u_d= \frac{1}{2} (u_F+u_R) = 1- p
-\frac{1}{2} \rho_R-\frac{1}{2} p/\rho_F.
\end{eqnarray}
A cluster maintains its integrity only
if its internal bulk density $\rho_c$ is stationary. The number of
particles in the cluster, $N_c$, is determined by nonlocal hops
at its two ends, and is thus on average constant in time,
$\partial  N_c /\partial  t \simeq  -p +p =0$. A stationary
internal density $\rho_c$ requires therefore that the average
cluster length $\xi= x_F-x_R$ be invariant.
Equations~(\ref{clusterfront}) and~(\ref{freeclusterrear}) then relates
the front and rear densities as
\begin{eqnarray}
\label{freeclusterlength}
\rho_R\cdot \rho_F=p.
\end{eqnarray}
Inserting reasonable values for the densities
yield immediately that $u_d$ changes sign as $p$ increases, and does so
before the MF estimate of $p_c$.
For example, $\rho_F=1/2$ implies that the drift velocity changes
sign at $p_c=1/3$;
and also that $\rho_R=2/3$, large enough to validate the cluster
concept self-consistently.

This turn-about of the drift
velocity of the clusters explains the first-order nature of the
MC-ER transition. Any initial configuration decays (after forming
clusters) to an empty road for $u_d <0$, because all bulk clusters
travel backward and fall back into the entry point reservoir.
This change in drift velocity is clearly visible in the density
profile time evolutions of Fig.~\ref{fig.clustering}.

Emboldened by this success, we dare to push the self-consistent
cluster analysis further. We need to relate the above equations
to the free cluster current $J_c$ and bulk
density $\rho_c$ to obtain a closed set of equations.
To determine the free cluster properties throughout the C and MC
phases, and also estimate the value of $p_c$.

Visualize each cluster as a separate conventional ASEP, i.e.,
the absence of significant numbers of internal nonlocal hopping events
due to the high local internal density. In conventional ASEP
the density probability distribution deep inside
the bulk is spatially uncorrelated. It is safe to presume the same is true
inside the clusters. This yields
\begin{eqnarray}
\label{clusterbulkcurrent} J_c=(1-r v_c)v_c,
\end{eqnarray}
with $r=1-p$. The cluster acts like an individual ASEP with open
boundary conditions in a somewhat unusual set-up: It has a
fluctuating `lattice size' $\xi$ and a fixed injected current $p$,
instead of a fixed lattice size and a fluctuating injection
current (controlled by a fugacity like $\alpha$). The total
current inside a cluster is equal to the injection current plus
the contribution of the center of mass drift as
\begin{eqnarray}
\label{freeclustercurrentdrift} J_c = p + \rho_c u_d.
\end{eqnarray}
$J_c$ and $\rho_c$ should not be confused with the global current
and density. They only coincide with those at the transition point $u_d=0$.
Eq.~(\ref{freeclustercurrentdrift}) reproduces correctly the exact
result that $J=p_c$ at the transition.

Equations~(\ref{freeclusterdrift})-(\ref{freeclustercurrentdrift})
provide still only 4 conditions between 5 variables:
the cluster current $J_c$, the drift velocity $u_d$,
and the rear, bulk, and front densities ($\rho_R$, $\rho_c$, and $\rho_F$).
Typically (at a given value for $p$), a set of possible solutions can be found.
Let's parameterize them by their values for $J_c$.
We use the maximal current principle to close the equations and
select the correct value of $J_c$ from the set.
The justification of this is again based on  conventional ASEP,
where the exact matrix method solution as well as the MF
approximation have this property~\cite{Derrida}. A secondary feature of the conventional
ASEP is also preserved, that the stationary state density profile is always
uniformly decreasing, $\rho_R\geq \rho_c\geq \rho_F$.

This procedure leads to two distinct type free cluster solutions as
function of $p$: a `front-limited' (F) state (with $\rho_c=\rho_F$) at $p>p_s$;
and a `maximum-current' (MC) state (with $\delta J/\delta v_c=0$) at $p<p_s$.
The crossover takes place at
$p_s=\frac{1}{4}(3-\sqrt5)=0.191$.

The MC free cluster solution at $p<p_s$ resembles
very closely and smoothly connects to the $p=0$ global MC state.
It has internal vacancy density $v_c=\frac{1}{2r}$, cluster current
$J_c=\frac{1}{4r}$, drift velocity $u_d=r-\frac{1}{2}$, and edge
densities $\rho_F=2p$ and $\rho_R=\frac{1}{2}$.
However, we cannot take the MC free cluster solution seriously
at small $p$, because the internal cluster density becomes too small for
the cluster concept to remain meaningful. Instead, we interpret
$p_s$ as marking the onset of crossover towards the non-clustered
MC stationary state at $p=0$.

The F solution with $\rho_c=\rho_F$,
maximizes the internal current inside the cluster for $p>p_s$.
The cluster density is equal to
\begin{eqnarray}
\label{Fstatebulkdensity}
\rho_c= \frac{1}{2(1-p)}\Big[\sqrt{4p-3p^2}-p\Big].
\end{eqnarray}
The rear density, $\rho_R=p/\rho_c$, is larger than
$\rho_c=\rho_F$ in the entire interval $p_s<p<p_c$, reflecting
a uniformly decreasing  density profile (required for internal stability).

These F solutions tell us that inside the MC phase (bulk-in-control phase)
of Fig.~\ref{fig.palpha}, near $p_c$, the configurations are
clustered and that each of these free clusters is limited
and controlled by its front.
Inserting $\rho_c=\rho_F$ from Eq.~(\ref{Fstatebulkdensity}) into
Eq.~(\ref{clusterfront}) yields
$u_d=0$ at $p_c=1/3$. This estimate for the first-order transition
line is remarkably close to our numerical results, $p_c\simeq 0.30$.

\section{Mother cluster analysis}
\label{mother}

We now turn our attention to the the critical line
between the C and ER phases. This transition is dominated
by the fluctuations near $x=1$.
Therefore, we extend the self-consistent
cluster analysis to the mother cluster, i.e., the cluster attached to
the entry point reservoir.  This analysis explains the
scaling properties of the second-order transition, as determined
numerically and presented in the following sections.

The mother cluster is presumed to be a mesoscopic object, connected
to the entry side reservoir. It governs the bulk current in the C
phase, while being controlled itself by the reservoir.
The mother cluster acts like a $p=0$ ASEP with a fluctuating front,
i.e., with a fluctuating length.
The injection current from the reservoir in its rear is
equal to $J_c= \alpha v_1$, with $v_x=1-\rho_x$ as before.
The bulk density of the mother cluster can be presumed to be
uncorrelated again, Eq.(\ref{clusterbulkcurrent}).
The evolution of its front at $x_F$ obeys Eq.~(\ref{clusterfront}).
An amount $p$ of the current through the cluster flows away at the front of the
cluster by nonlocal hops, and the remainder extends the cluster forward:
\begin{eqnarray}
\label{motherfrontcurrent}
J_c= p+ \rho_c u_F= p+\rho_c\big[(1-p)-p/\rho_F\big].
\end{eqnarray}
This set of equations is still incomplete. It ignores the
probability $P_0(x)$ that all sites to the left of a specific site
$x$ are empty, see Eq.~(\ref{bulkcurrent}). In the free cluster
analysis this played no role ($P_0(x)=0$) because another particle can
always be found to the left of the free cluster as long as the
mother cluster exists and remains mesoscopic.
The mother cluster current is very sensitive to the density profile
near the reservoir, and thus to $P_0(x)$.

The mother cluster density profile can be presumed (self-consistently)
to be sufficiently uniform that we can apply mean field theory internally,
\begin{eqnarray}
\label{motherclusteriteration}
v_{x+1}&=&J_c/[1-rv_x- p P_0(x)], \\
P_0(x)&=&v_x P_0(x-1).
\end{eqnarray}
Compare this with Eq.~(\ref{bulkcurrent}).
MF theory for conventional ASEP gives erroneous
power-law and exponential density profile tails.
Similarly, the following analysis can be at best
only qualitatively correct.

Solutions are found by solving the above equations iteratively
and by finding the value of $J_c$
that satisfies the boundary conditions at both the rear and front.
It is important to realize that the solution is
unique for `finite' cluster lengths $\xi_c$.
The mother clusters are finite in length,
but we analyze them as if they are mesoscopic
and effectively infinitely long.
For infinite cluster lengths, the equations
typically allow a range of $J_c$ with possible solutions.
The maximum current solution is the correct one, because it
coincides with the $\xi_c\to \infty$ limit of
the unique finite length solution.

At $\xi_c=\infty$, the construction of a solution involves
the matching of two iteration processes:
the forward iteration, $x\to x+1$, starting from the rear of the
cluster and the backward iteration,
$x\to x-1$, starting from the front of the cluster.
It is useful to sketch briefly some of the details of this.
The iteration process has two bulk densities as possible fixed points,
determined by the roots of the quadratic
equation $J_c=v_c(1-rv_c)$ (using that $P_0\to 0$ in the bulk of the mother
cluster). The low density fixed point $v_c^{(l)}$ is unstable and
the high density fixed point $v_c^{(h)}$ is stable in the forward
iteration process. They reverse roles in the backward iteration process.

\begin{figure}[t]
\includegraphics*[width=0.7\columnwidth, angle=0] {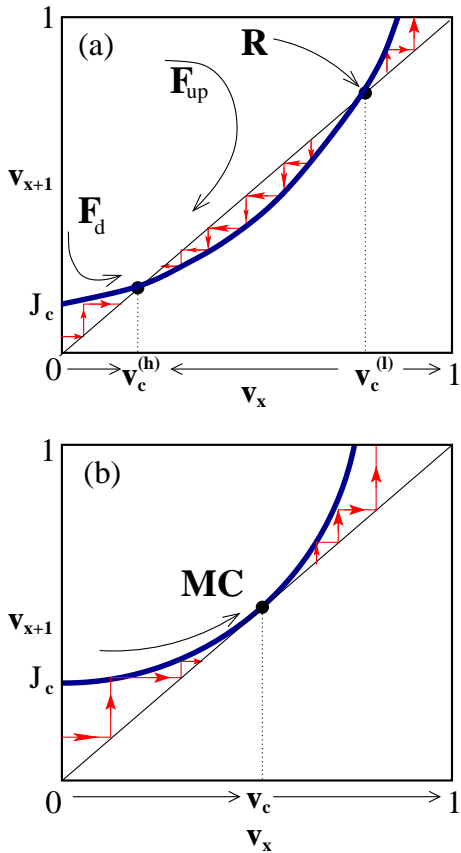}
\caption{(Color online) Forward iteration processes. }
\label{fig.iteration}
\end{figure}

For a given value of $J_c$, the forward iteration process starts
at site $x=1$ with density $v_1=J_c/\alpha$, and typically iterates
towards $v_c^{(h)}$ (see Fig.~\ref{fig.iteration}).
The backward iteration process starts at site
$x=x_F$ with density  $v_F$ prescribed  by Eq.~(\ref{motherfrontcurrent}),
and typically converges towards $v_c^{(l)}$. Those solutions do not match,
unless the current is raised to the value where the two fixed points coincide.
This matching is equivalent to (and the origin of) the maximal current principle.
This type of solution, where the two fixed points merge,
applies to the bulk controlled MC state. It exists
when the starting densities of the forward and backward
iteration processes are located inside the attraction domain of
the $v_c^{(h)}$ and $v_c^{(l)}$ fixed points, respectively.

\begin{figure}[t]
\includegraphics*[width=0.7\columnwidth, angle=0]{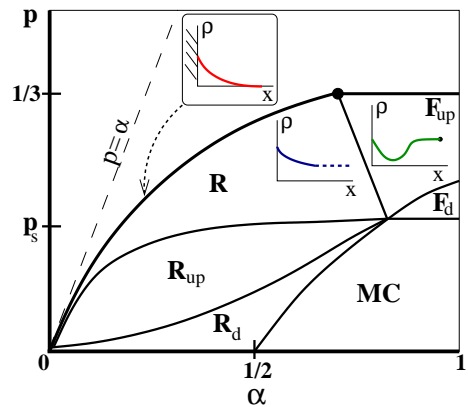}
\caption{Mother cluster states in the $\alpha$-$p$ phase diagram.}
\label{fig.mothercluster}
\end{figure}

This fails in the front and rear controlled, F- and R-type states.
In the F states, the starting point of the backward
iteration process, the value $v_F$, crosses over to the unstable
side $v_F<v_c^{(h)}$ of the $v_c^{(h)}$ fixed point,
while we raise the current. In that case $J_c$ must be chosen such that
those two densities match, $v_F=v_c^{(h)}$. The density profile at the
front is thus flat. The rear profile is typically able to iterate forward
towards this fixed point $v_c^{(h)}$, and thus match the backward iteration.
(This solution still maximizes the current, in the boundary limited
sense.)

In the rear-controlled (R) states, the roles of the forward and
backward iteration processes are reversed, i.e.,
the forward iteration process crosses over to the wrong side of
the $v_c^{(l)}$ fixed point. This requires more care than
the F state, because the initial value
$v_1=J/\alpha$ needs to be iterated forward into the region
$x\simeq l_p$ where $P_0(x)$ becomes negligible, before it becomes
clear where this profile `lands' compared to the low density fixed
point $v_c^{(l)}$. The latter is the reason why we need to bother
with the iteration process for the mother cluster, while for the
free clusters it suffices to write down the boundary and bulk
conditions, and use the maximal current principle formulation
of the matching process.

Figure~\ref{fig.mothercluster} shows the mother cluster states
as a function of $\alpha$ and $p$. The MC (bulk-controlled) and F
(front-controlled) states are basically identical to the
free cluster states discussed in the earlier section. This
confirms the stability of the MC clustered phase at large
$\alpha$. In these states the front or bulk of the mother cluster
is in control, setting $J_c$, and at the rear the reservoir is
able to supply the requested current without letting the mother
cluster detach. The F state can be divided into two states, F-up
and F-down, distinguishing between
whether the forward iteration process approaches $v_c^{(h)}$
from below or above (see Figs.~\ref{fig.iteration}
and \ref{fig.mothercluster}).
In the free cluster analysis only F-down is present.
Due to the presence of $P_0(x)$, the mother cluster is not
required to be uniformly decreasing any more.

These F and MC solutions describe mesoscopic mother clusters.
The growth velocity (drift velocity of the front)
$u_F$ is positive. Strictly speaking the mother cluster
could keep growing until it fills the entire chain,
but that would lead to the uniform MF-type
non-clustered state along the entire chain. In reality, density
fluctuations make the mother cluster break apart regularly,
emitting free clusters.

The transition into the R states at small $\alpha$ takes place
when the front (F-state) or bulk (MC-state)
imposed value of $J_c$ can not be satisfied at small $x$ anymore.
From there on, the rear takes over as the limiting factor,
and the front end of the profile is required to follow instead.
This is the origin of the starvation process responsible for the
second-order transition line.

The mother cluster front can only follow the rear at small $p$,
denoted in Fig.~\ref{fig.mothercluster} as the R-up and R-down states.
The current $J_c$ must be tuned such that the forward iteration process
lands exactly onto the low density fixed point $v_c^{(l)}$. In the
R-up and R-down states, the backward iteration process, starts
at a $v_F$ inside the attraction basin of the low density fixed point,
and therefore the backward iteration process is able to converge.

The R-up state has an increasing density profile in the front, where
$P_0=0$, and is therefore dynamically unstable. Moreover, the bulk
densities in both the R-up and R-down states are very small, just
as in the MC state, and self-inconsistent with the cluster concept.
The mother cluster density profile starts high at $x=1$,
but decays to the small unstable fixed point value beyond $x=l_p$.
Such clusters must be unstable to internal clustering fluctuations
and break-up regularly into a small $l_p$ sized object.

Everywhere near the second-order transition line,
in the area marked as R in Fig.~\ref{fig.mothercluster},
the backward iteration process fails because
$v_F$ starts at a $v_F<v_c^{(h)}$ outside the attraction basin of
$v_c^{(l)}$. The forward and backward iterations can not be matched anymore.
A true R-type mother cluster solution seizes to exist.
The front of the mother cluster can not be
in a locally stable stationary type forward creeping state anymore.

This inability to construct a stable R cluster front near the
second-order transition line in Fig.~\ref{fig.palpha},
leaves no other options than to toss out
the mother cluster front conditions,
Eq.~(\ref{motherfrontcurrent}),
and redefine the mother cluster as a quasi-stationary R-type object
(typically short, except very close to the second order transition line)
defined by the half-space forward iteration process only,
with $J_c$ the precise value of the cluster current that let
$v_1=\alpha/J_c$ iterate into $v_c^{(l)}$.
The mother cluster size is set by the length scale $l_p$ where
$P_0(x)\to 0$ and the density becomes small.
$l_p$ is now the onset of a turbulent low density region
where the formation of new free clusters takes place,
continuously without any ``local stationary state order" or stability.

Next, we face the problem that we lost our (maximum current) justification
for why $J_c$ must settle on that special value
where the forward iteration process
converges into the unstable fixed point. Nevertheless, it is arguably the
dynamically stable choice in the following sense. Consider the half-space
solutions with smaller values of $J_c$. Their forward iteration process
converges into the stable high density fixed point. Those clusters would have
a pronounced dip in their density profile near $l_p$
(when the iteration process sails near the $v_c^{(l)}$ fixed point)
before increasing towards the large density value of the $v_c^{(h)}$ fixed point.
Such structures are unstable and tend to shrink. They resemble mother
clusters in the process of pinching-off a free cluster.
The breaking-up event shrinks the mother cluster towards $l_p$.
Moreover, the break-up also increases the current of the combined two objects.
Half-space solutions with values of $J_c$ larger than the special choice,
make the iteration process crash almost immediately through zero density,
and can be interpreted as representing very short mother clusters.
Those are also unstable and tend to grow. The empty space in front of
such a short cluster can carry away still only a current of order $p$,
while $J_c$ has increased. Moreover, it is less successful in breaking-up
because $P_0(x)$ is not small at its front,
and therefore the free cluster immediately in front of it
is subject to a fluctuating injection current, larger on average than $p$,
and thus with a significantly smaller or even negative drift velocity.

The C-MC phase boundary is represented in the mother cluster analysis
as the threshold, see  Fig.~\ref{fig.mothercluster}, between the R
states and the MC and F states. These reproduce the true location
(from our numerical simulations) not too well.
It has a kink instead of being smooth.
The reason for this is likely that the mother cluster is very short
in the R-states and badly approximated by the mesoscopic description.

The C-ER critical line is represented in the mother cluster analysis
as the line where the R-state solution iterates to zero density.
This estimate virtually coincides with the true location determined
from our numerical simulations.
This remarkable accuracy is likely linked to
the fact that the mother cluster length, of order $l_p$,
diverges at the transition, and the mesoscopic approach becomes valid.

\begin{figure}[t]
\includegraphics*[width=0.7\columnwidth, angle=0]{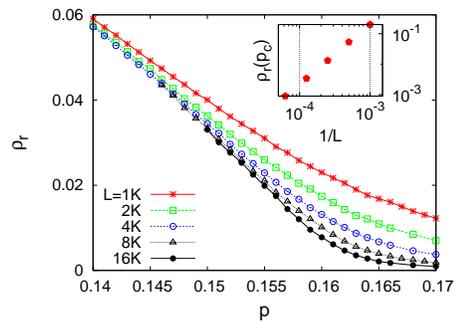}
\caption{(Color online) The road density $\rho_r$ versus $p$ at
$\alpha=0.2$ for various $L$. The inset illustrates
the finite-size behavior of $\rho_r$ at the C-ER
transition point in a log-log plot.}
\label{fig-rho-a2}
\end{figure}

The crossover from the horizontal first-order segment of the
transition line to the curved second-order part represents
a critical endpoint. It is located at exactly $p_c=1/3$ and
$\alpha=2/3$ in this analysis, and is also close to the true location.
Our numerical simulations locate it at $p_c\simeq 0.300(2)$
and $\alpha\simeq 0.64(1)$. The scaling properties of the critical endpoint
appear numerically not special, and consistent with the notion,
emerging from the cluster analysis, that the origins of the first- and second-order
transitions (starvation versus drift-velocity turnabout) are distinct
and unrelated. Their scaling behaviors are thus simply superimposed onto each
other at the critical endpoint.

Within our numerical accuracy this critical endpoint coincides with the endpoint of
the C-MC boundary. There is no a priori reason to expect this coincidence,
but in the mother-cluster approach it arises naturally.

\section{Numerical results at criticality}
\label{numerical}

\begin{figure}[t]
\includegraphics*[width=0.7\columnwidth, angle=0]{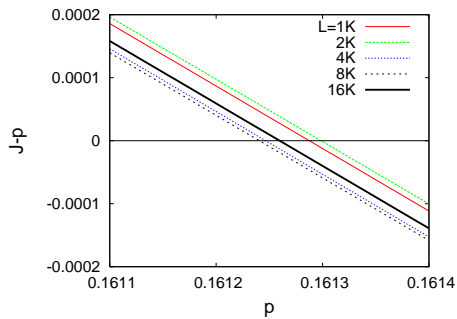}
\caption{(Color online) Details of $J-p$ versus $p$
as a function of $p$ at $\alpha=0.2$ for various $L$ show that
finite-size corrections are very small and $J(p_c)=p_c=0.16125(5)$.
Compare with Fig.~\ref{fig.current-p-p}.}
\label{fig-current-a2}
\end{figure}

In our numerical simulations, we average over $10^4\sim 10^5$
independent runs up to $t=10^5$ Monte Carlo steps
for non-stationary state features. For the stationary state properties
we simulate system sizes $L=500\times 2^n$ with $n=1,2\cdots, 5$
and discard all data before $t\simeq L^2$, or more, to allow the
system to reach the stationary state. We collect data up to
$t=10^8\sim 10^9$ Monte Carlo steps and average over $10\sim 200$
independent runs. We monitor: the road density $\rho_r=\langle
n_r\rangle$ ($n_r=N_r/L$); the density fluctuations,
$\chi_2 =L
(\langle n_r^2\rangle -\langle n_r \rangle^2)$; the Binder
cumulant $U_4=1- \langle n_r^4 \rangle/(3\langle n_r^2
\rangle^2)$; and the current $J$. The results are shown in
Figs.~\ref{fig-rho-a2} and~\ref{fig-current-a2}.

The scaling properties of the C-ER transition are determined
numerically, along several cuts though the transition line.
It suffices to present here the details
along one representative example, i.e., only the $\alpha=0.2$ line.
The critical point $p_c$ along this line can be determined
by standard methods, from the so-called crossing points
or the maximum points in $\chi_2$ as well as the crossing points in $U_4$.
From these we estimate $p_c=0.161(1)$.
As mentioned before, the current, see Fig.~\ref{fig-current-a2},
must be equal $J=p$ at the transition line. This is very well
satisfied compared to the above estimate for $p_c$. Turning this
around, requiring $J(p_c)=p_c$,
improves the estimate to $p_c=0.16125(5)$.

\begin{figure}[b]
\includegraphics*[width=0.7\columnwidth, angle=0]{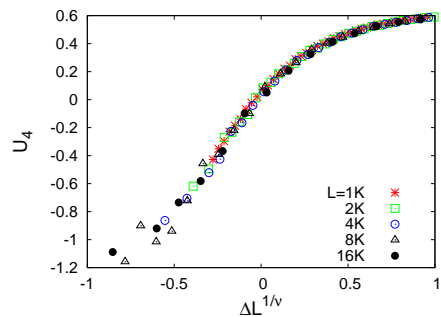}
\caption{(Color online) Scaling collapse of the Binder cumulants at
$\alpha=0.2$
with $p_c=0.16125$ and $1/\nu=0.50$.}
\label{fig-BC-collapse-a2}
\end{figure}

\begin{figure}[t]
\includegraphics*[width=0.7\columnwidth, angle=0]{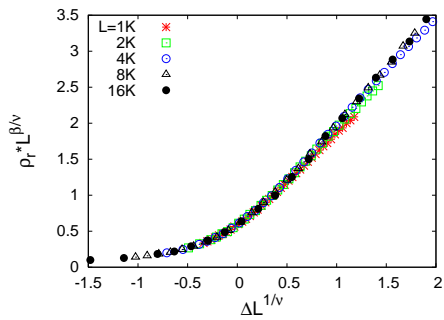}
\caption{(Color online) Scaling collapse of $\rho_r$ at $\alpha=0.2$ with
$p_c=0.16125$, $\beta/\nu=0.48$, and $1/\nu=0.53$.}
\label{fig-rho-collapse-a2}
\end{figure}

\begin{figure}[b]
\includegraphics*[width=0.7\columnwidth, angle=0]{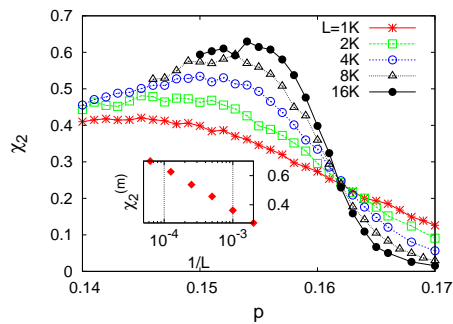}
\caption{(Color online) The density fluctuations, $\chi_2$, versus $p$ at
$\alpha=0.2$ for various $L$. The inset shows the finite-size
behavior of the maximum values, $\chi_2^{\rm (m)}$, in
a semi-log plot.}
\label{fig-chi_2-a2}
\end{figure}

The road density $\rho_r$ vanishes on approach of the transition point $p_c$
from below in the thermodynamic limit,
as $\rho_r\sim \Delta^\beta$ with $\Delta=p_c-p$.
The density fluctuation $\chi_2$ diverges at $p_c$ from below as
$\chi_2 \sim \Delta^{-\gamma}$ but becomes zero in the ER phase
($p>p_c$). The Binder cumulant $U_4$ converges to $2/3$ in the C
phase (consistent with a Gaussian distribution) and to $-1$ in the
ER phase (indicating an exponential distribution). $U_4$ converges
very fast, while $\chi_2^{\rm (m)}$ shows strong finite-size effects.

The values of the critical exponents follow by exploiting
conventional finite-size scaling (FSS) theory
\begin{eqnarray}
\rho_r &=& L^{-\beta/\nu} f(\Delta L^{1/\nu}),\nonumber\\
\chi_2 &=& L^{\gamma/\nu} g(\Delta L^{1/\nu}),\\
U_4 &=&  h(\Delta L^{1/\nu}),\nonumber \label{eq-scaling}
\end{eqnarray}
where $\nu$ is the correlation length exponent and $f$, $g$, and
$h$ are scaling functions. Elementary scaling laws yield
hyperscaling relations such as $(2\beta+\gamma)/\nu=1$.

$U_4$ shows the least finite-size effects. We estimate the value
of $\nu$ by collapsing the $U_4$ data, see
Fig.~\ref{fig-BC-collapse-a2}, as $1/\nu=0.50(2)$. Collapsing the
road density data yields, see Fig.~\ref{fig-rho-collapse-a2},
$\beta/\nu=0.48(2)$ and $1/\nu=0.53(3)$. By plotting the maximum
value of $\chi$ against $L$, we can estimate the value of
$\gamma/\nu$. A simple power law fitting yields
$\gamma/\nu=0.17(3)$, but the data is also consistent with a
logarithmic fit, i.e., $\chi_2^{\rm (m)}\sim \log L$,
see Fig.~\ref{fig-chi_2-a2}. Hyperscaling is satisfied only when
this logarithmic divergence is correct, so we are inclined to put
more weight on the logarithmic fitting.

The value of the Binder cumulant $U_4(p_c)$ is expected to be
universal at the transition. Our numerical estimate,
$U_4(p_c)=0.08(3)$, is very close to zero, which may reflect the
half-Gaussian like shape of $P(N_r,L)$ at criticality, see
Fig.~\ref{fig.histoC}.

\section{Mother and free cluster separation}
\label{separate}

The numerical data point to rather simple values of the critical
exponents at the C-ER transition, $\beta=1$ and $\nu=2$. The
R-state mother cluster properties explain why the scaling is
that simple. The first and most crucial observation is the lack of any feature in the
current $J(p,\alpha)$ across the critical line in our numerical data
Figs.~\ref{fig.current-p-p} and \ref{fig-current-a2}.
The global current appears to be analytic,
and totally oblivious of the presence of the C-ER
transition, moving linearly through the critical value $J=p$,
as $J-p\simeq A(p-p_c)$.

The mother cluster sets the global average current.
Its offspring, the free clusters, cannot
influence this, unable to talk back being distinct
objects with positive drift velocity. The mother cluster is an
independently fluctuating object, and the rest of the road is
slaved to it. At the C-ER transition, the mother
becomes unable to create free clusters at a sufficient rate to
sustain a non-zero
average bulk density $\rho_c$, but is itself unaffected by
what is going-on in the bulk.

The global average current $J(\alpha,p)$ is analytic
across the C-ER transition,
because the mother cluster current is analytic, and
the average  bulk current is equal to it, because
(on average) the mother cluster remains attached to the
reservoir, and (on average) it does not grow in length
(due to the breaking-off of free clusters).

The linear vanishing of
the bulk density with exponent $\beta=1$ is
a direct consequence of this
analyticity of $J(\alpha,p)$.
They are related by the simple equation
\begin{eqnarray}
\label{bulkdensity1}
J(\alpha,p)&=&p(1-\rho_r) +(p+\rho_c^{(f)} u_d^{(f)})\rho_r \\
\label{bulkdensity2}
\rho_r&=&\frac{J-p}{\rho_c^{(f)} u_d^{(f)}}
\end{eqnarray}
with $\rho_c^{(f)}$ the internal free cluster bulk density  and
$u_d^{(f)}$ their drift velocity
(see also Eq.~(\ref{freeclustercurrentdrift})).
Consider a specific bond. The first term
on the right hand side of Eq.~(\ref{bulkdensity1})
represents the current through this bond from isolated nonlocally
hopping particles when this bond  is not part of a free cluster.
The second term is the current through the bond when it is part of
a free cluster. The probability for the bond to be part of
a free cluster is equal to $\rho_r$. The free cluster parameters
$\rho_c^{(f)}$ and $u_d^{(f)}$ are analytic functions of $p$ only and set
self-consistently by the free cluster equations.
All functions on the right hand side of Eq.(\ref{bulkdensity2}) are analytic,
therefore the bulk density vanishes linearly, with $\beta=1$.

The C-ER transition is a bulk phenomenon induced by the mother cluster.
It lacks therefore independent critical bulk fluctuations.
This explains why $\nu=2$. The free cluster generating break-ups
in the turbulent region near $x=l_p$
behave like random uncorrelated events; i.e., the fluctuations in the
number of particles injected by the mother cluster
into the bulk behave like uncorrelated noise.
In the C phase, those fluctuations are screened and absorbed
by the presence and formation of free clusters.
On approach of the C-ER transition,
free clusters become rare and at the transition
they vanish altogether. There, the random break-up fluctuations
travel like a pattern across the chain.
The time of flight is proportional to the system size.
Therefore, the bulk density represents (biased) random noise
averaged over a time proportional to $L$,
and thus scales as $\rho_r\sim L^{-1/2}$ at the C-ER transition.

Two more variables in our numerical data express
and confirm the same uncorrelated fluctuations.
At the C-ER transition, the density profile (no figure included)
decays to zero smoothly near the reservoir, as a power-law,
$\rho_r(x)\sim x^{-\delta_R}$ with $\delta_R\approx 1/2$,
consistent with random noise. In the C phase,
the distribution of the number of particles on
the road, $P(N_r)$, is a Gaussian centered
around the bulk density $\rho_r$. Towards the C-ER transition,
the maximum shifts towards $N_r\approx {\cal O}(1)$ and
at the transition decays as a half Gaussian
(see Fig.~\ref{fig.histoC}) consistent with uncorrelated noise.
In the ER phase, $\rho_r(x)$ and $P(N_r)$ decay exponentially.

\section{Conclusions}
\label{ending}

In this paper, we studied how the inclusion of nonlocal hopping events
affects the stationary state of ASEP. Clusters develop, only mesoscopic ones.
There is no macroscopic queuing transition. Instead,
there is a phase transition towards an ``empty-road phase",
which is first-order from the MC clustered phase
(where the bulk is in control) and
second-order from the C-type clustered phase
(where the entry reservoir controls the stationary state).
The first-order MC-ER transition is induced by a reversal
of the group velocity of the free clusters.
The C-ER transition has rather simple scaling properties,
reflecting the mother cluster properties near the entry reservoir
and that the bulk (free clusters) do not interact
back to the mother cluster.

One of the remaining issues is the nature of the dynamic scaling.
Generically, fluctuations in the stationary state broaden in time as
$l\sim t^{1/z}$. The dynamic scaling in the clustered C and MC phases, is
to be expected to remain in the KPZ universality class, with
$z=3/2$. Within each cluster length scale we expect $z=3/2$
because each cluster acts like a simple ASEP in its own right. At
larger length scales, the process resembles a driven zero-range
process with a preferred direction and non-zero group velocity,
which should also have $z=3/2$ scaling. In our simulations, we did
not focus on the dynamic scaling properties of the C ad MC
phases, but exploratory numerical results for $z$ are
consistent with $z=3/2$. The dynamic scaling at the C-ER transition
is expected to remain simple as well because that transition does
not involve novel intrinsic bulk fluctuations.

Clustering and its ramifications are surely the main message of
our study. The conventional ASEP has a uniform stationary state with
rather trivial scaling properties; but is unstable towards clustering
and queueing. Such clustered states, even when involving only mesoscopic
clusters, communicate badly with each other, and have a hard time
developing novel type fluctuations. Phase transitions,
induced and controlled by those same clusters,
therefore have typically rather simple scaling properties,
like at the C-ER and MC-ER phase boundaries in our process.

It will be interesting to test our self-consistent cluster
approximations to other processes with clustering in cases where
the drift velocity of the clusters is non-zero; for example, a
variant of the two species process by Arndt {\it et
al}.~\cite{AHR} with different numbers of opposite moving
particles. 

\section*{Acknowledgments}
We like to thank D.~Mukamel, G.M.~Sch{\"u}tz, and R.P.K.~Zia for
useful discussions. This work was supported by the National
Science Foundation under Grant No. DMR-0341341
and by the BK21 project.

\end{document}